\documentclass[reprint,amsmath,amssymb,aps,prb,superscriptaddress]{revtex4-1}

\usepackage{textcomp,pxfonts}
\usepackage{amsmath}
\usepackage{graphicx}
\usepackage{multirow}

\usepackage{color}
\begin{document}



\title{Anharmonic lattice dynamics of Ag$_2$O studied by inelastic neutron scattering and first principles molecular dynamics simulations}



\author{Tian Lan} 
\email[]{tianlan@caltech.edu}
\affiliation{\mbox{Department of Applied Physics and Materials Science, \nolinebreak California Institute of Technology, Pasadena, California 91125, USA}}
\author{Chen W. Li}
\affiliation{Oak Ridge National Laboratory, Oak Ridge, Tennessee 37831, USA}
\author{J. L. Niedziela}
\affiliation{Oak Ridge National Laboratory, Oak Ridge, Tennessee 37831, USA}
\author{Hillary Smith}
\affiliation{\mbox{Department of Applied Physics and Materials Science, \nolinebreak California Institute of Technology, Pasadena, California 91125, USA}}
\author{Douglas L. Abernathy}
\affiliation{Oak Ridge National Laboratory, Oak Ridge, Tennessee 37831, USA}
\author{George R. Rossman}
\affiliation{Division of Geological and Planetary Sciences, California Institute of Technology, Pasadena, California 91125, USA} 
\author{Brent Fultz}
\affiliation{\mbox{Department of Applied Physics and Materials Science, \nolinebreak California Institute of Technology, Pasadena, California 91125, USA}}



\date{\today}

\begin{abstract}

Inelastic neutron scattering measurements on silver oxide (Ag$_2$O) with the cuprite structure were performed at temperatures from 40 to 400\,K, and Fourier transform far-infrared spectra were measured from 100 to 300\,K. The measured phonon densities of states and the infrared spectra showed  unusually large energy shifts with temperature, and large linewidth broadenings. First principles molecular dynamics (MD) calculations were performed at various temperatures, successfully accounting for the negative thermal expansion (NTE) and local dynamics. 
Using the Fourier-transformed velocity autocorrelation method, the MD calculations reproduced the large anharmonic effects of Ag$_2$O, and were in excellent agreement with the neutron scattering data. The
quasiharmonic approximation (QHA) was less successful in accounting for
much of the phonon behavior.
The QHA could account for some of the NTE below 250 K, although not at higher temperatures.
Strong anharmonic effects were found for both phonons and for the NTE. 
The lifetime broadenings of Ag$_2$O were explained by anharmonic perturbation theory, which showed rich interactions between the Ag-dominated modes and the O-dominated modes in both up- and down-conversion processes. 

\end{abstract}




\maketitle

\section{Introduction}
\parskip=0.in

Silver oxide (Ag$_2$O) with the cuprite structure has attracted much interest after the discovery of its extraordinarily large negative thermal expansion (NTE),\cite{NTEa, NTEb} which exceeds $-1 \times 10^{-5}$ K$^{-1}$ 
and occurs over a wide range of temperature from 40\,K to its decomposition temperature near 500\,K.  
Besides its large NTE, Ag$_2$O is commonly used as a modifier in fast-ion conducting glasses and batteries, \cite{battery,ions} 
and its catalytic properties are also of interest. \cite{catalysis1, catalysis2}

\begin{figure}[t]
\includegraphics[width=0.6\columnwidth]{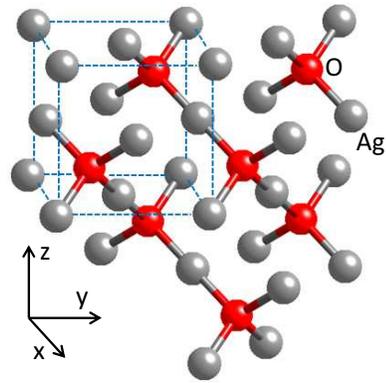}
\caption{Cuprite structure of Ag$_2$O, showing standard cubic fcc unit cell and O-Ag-O links
that pass between cubes.}
\label{fig:structure}
\end{figure}

In the cuprite structure of Ag$_2$O  shown in Fig. \ref{fig:structure},
the fcc Ag lattice is expanded by the presence of the O atoms,
which form an interpenetrating bcc lattice.
The O atoms occupy two tetrahedral sites of the  standard fcc unit cell of Ag atoms. 
Each O atom is linked to four O atoms through a bridging Ag atom,
placing the Ag atoms in linear O-Ag-O links with little transverse constraint.
A geometrical model of NTE considers tetrahedra of Ag$_4$O around each O atom
that bend at the Ag atoms linking the O atoms in adjacent tetrahedra.
Rigid-unit modes (RUMs) account for counteracting rotations of all such tetrahedra. \cite{RUMa, RUMb}
These RUMs tend to have low frequencies owing to the large mass of the unit, 
and hence are excited at low temperatures. 
Locally, the O-Ag bond length does not contract, but bending of the 
O-Ag-O links pulls the O atoms together,
leading to NTE. 
This model correlates the NTE and lattice dynamics.
Similar models seem to explain the large NTE of ZrW$_2$O$_8$ and other systems.

The RUM model has value even if the Ag$_4$O tetrahedra do not move as rigid units,
but the interpretation of NTE becomes less direct. 
A related concern is that modes involving bending of the O-Ag-O links may  be strongly anharmonic.
In a recent study on ScF$_3$, for example,   the NTE largely originated with the bending of linear Sc-F-Sc links. \cite{ScF3}
Frozen phonon calculations showed that the displacement of F atoms in these modes followed a nearly
quartic potential, and the low mass of F made it possible to approximate the problem as 
independent local quartic oscillators.
The heavy mass of the Ag atoms in the O-Ag-O links implies that different Ag atoms
will move cooperatively, and delocalized anharmonic oscillators are  challenging to understand.

Recent measurements by
high-resolution x-ray diffractometry and extended x-ray absorption fine structure spectrometry (EXAFS) 
 showed large deformations of the
Ag$_4$O tetrahedra. \cite{exafsa, exafsb, chapman, exafsc} 
Although these tetrahedra are not distorted by a pure RUM, 
the simultaneous excitation of 
other modes  makes it unrealistic to view the 
dynamics as motions of rigid framework units. 
Measurements by EXAFS also showed that the average Ag-O nearest-neighbor distance
expands slightly upon heating, but the Ag-Ag next-nearest neighbor distance
contracts approximately as expected from the bulk NTE. 

 
Anharmonic  phonon behavior is  known to be important 
for  the thermodynamics and the thermal conductivity of materials at elevated temperatures. It is also important for the thermodynamic stability of phases. \cite{anhld1, anhld2} 
Anharmonic phonon behavior is sometimes associated with NTE, but 
such relationships are not well understood, and helped motivate the present study.

Inelastic neutron scattering is a powerful 
method to measure phonon dynamics, allowing accurate measurements of vibrational entropy. \cite{FultzReviewArticle2010}
Additionally, phonon energy broadening can be measured,
allowing further assessment of how anharmonic effects originate from the non-quadratic 
parts of the interatomic potential. 
A recent inelastic neutron scattering experiment on  Ag$_2$O with the cuprite structure
showed  phonon softening (reduction in energy) with temperature. \cite{inelasticneutron} 
The authors interpreted this result with 
a quasiharmonic model, where they calculated harmonic phonons for 
reduced volumes of the structure and obtained a negative Gr\"{u}neissen parameter. 
These measurements were performed on the neutron energy gain side of the
elastic line, restricting measurements to temperatures  above 150\,K, 
and the available energy range of  20\,meV allowed about a quarter of the Ag$_2$O phonon spectrum to be measured.

Lattice dynamics calculations, based  on either classical force fields or density functional theory (DFT), 
have been used to study materials with the cuprite structure. \cite{inelasticneutron,vibration,forcefield}
All these calculations were performed with the quasiharmonic approximation (QHA), 
where the interatomic forces and phonon frequencies changed with volume, 
but all phonons were assumed to be harmonic normal modes with infinite lifetimes. 
This QHA ignores interactions of phonons at finite temperatures through
the cubic or quartic parts of the interatomic potential, but these
interactions are essential to  explicit phonon anharmonicity.  
Although the QHA calculation accounted for the NTE behavior
in ZrW$_2$O$_8$,\cite{quasi} for  Ag$_2$O with the cuprite structure, the
QHA   was 
only partly successful. 
Molecular dynamics (MD) simulations should be reliable for
calculating phonon spectra in
strongly anharmonic systems ,\cite{autocorrb, autocorrc, autocorrd}
even when the QHA fails. 
To our knowledge, no MD investigation has yet been performed on  Ag$_2$O with the cuprite structure.   

To study  phonon anharmonicity in  Ag$_2$O, 
and its possible relationship to NTE,
we performed 
temperature-dependent inelastic neutron scattering experiments 
at temperatures from 40 to 400\,K to obtain the phonon density of states (DOS). 
(At  temperatures below 40\,K a first-order phase transition occurs, 
giving a temperature-dependent fraction of a second phase with different phonon properties \cite{NTEb,OldCalorimetry}.)
Fourier transform infrared spectrometry at cryogenic temperatures
was also used to measure the frequencies and lineshapes of phonons at the $\Gamma$-point
of the Brillouin zone.
First-principles ab-initio MD simulations were performed, and 
by Fourier transforming the velocity autocorrelation function,
the large temperature-dependent phonon 
anharmonicity was reproduced accurately.
An independent calculation of anharmonic phonon interaction channels 
was performed with interacting phonon perturbation theory, and 
semiquantitatively explained anharmonicites of the different phonons.
Most of the phonons have many channels for decay and are highly anharmonic.
Although the QHA is capable of predicting about half of the NTE at low temperatures, 
part of this NTE is associated with anharmonicity, 
and most of the NTE above 250\,K originates with anharmonic interactions
between Ag-dominated and O-dominated phonon modes.

\section{Experiments}

\subsection{Inelastic Neutron Scattering}
Inelastic neutron scattering measurements were performed with the wide 
angular-range chopper spectrometer, ARCS, \cite{arcs} at the Spallation Neutron Source at Oak Ridge National Laboratory. 
Powder samples of  Ag$_2$O with the cuprite structure of 99.99\% purity were loaded into an annular
volume between concentric aluminum cylinders 
with an outer diameter of 29 mm and an inner diameter of 27 mm, 
giving
about 5\% scattering of the incident neutron beam. 
The sample assembly was mounted in a bottom-loading closed cycle refrigerator
outfitted with a sapphire hot stage that can be controlled independently of the second stage of the cryostat.
Spectra were acquired with two  incident neutron energies of approximately 30 and 100\,meV. 
Measurements were performed at temperatures of 40, 100, 200, 300 and 400\,K,
each with approximately 1.6 $\times$ 10$^6$ neutron counts. 
Backgrounds with empty sample cans were measured at each temperature.
Ag has an
absorption cross section of 63 barns, but this was not be a problem with 
low background and  high neutron flux.

The raw data were rebinned into intensity $I$ as a
function of momentum transfer $Q$ and energy transfer $E$.
After deleting the elastic peak around zero energy,
neutron-weighted
phonon densities of states curves were calculated from
$I (Q,\,E)$ by subtracting the measured
background, and using an iterative procedure
to remove contributions from multiple scattering
and higher-order multiphonon processes. \cite{datareduce} 

\subsection{Fourier Transform Far-Infrared Spectrometer}

The far-infrared spectrometry measurements were performed with a Thermo-Nicolet Magna 860 FTIR spectrometer using a room-temperature deuterated triglycine sulfate detector and a solid substrate beam splitter.  The same Ag$_2$O powder was mixed with  polyethylene fine powder with a mass ratio of 1:19  and finely ground. The sample was compressed into a pellet of 1 mm  thickness, and mounted on a copper
cold finger of an evacuated cryostat filled with liquid nitrogen. The cryostat had polyethylene windows that were transparent in the far-infrared. Spectra were acquired at temperatures from 100 to 300\,K, and temperature was measured with  a thermocouple in direct contact with the sample pellet.  Backgrounds from a pure polyethylene pellet of the same size were measured at each temperature.

\subsection {Results} 

\begin{figure*}[t]
\includegraphics[width=1.8\columnwidth]{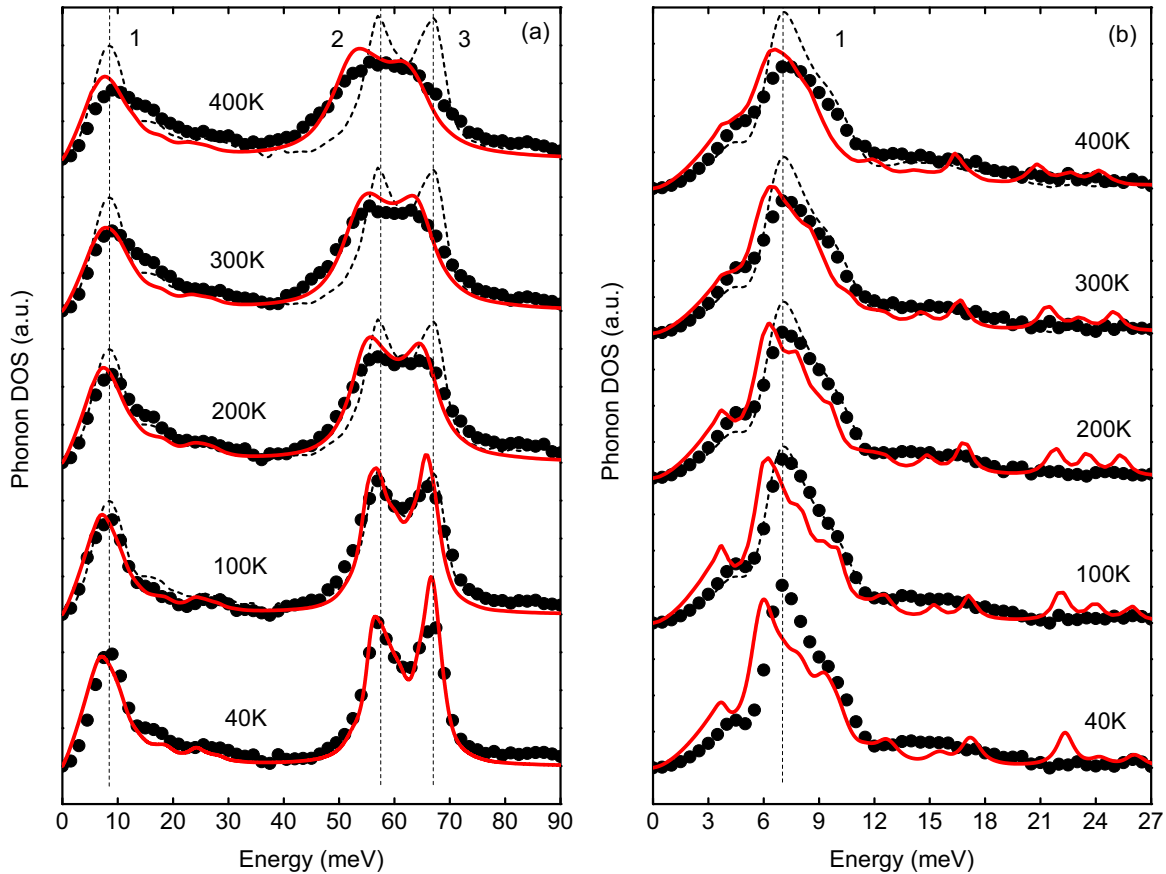}
\caption{Neutron weighted phonon DOS of  Ag$_2$O with the cuprite structure from ARCS experimental data (black dots) and MD simulations (red curves) at temperatures from 40 to 400\,K. The dashed spectrum corresponds to the 40\,K experimental result, 
shifted vertically for comparison at each temperature.  
Vertical dashed lines are aligned to the major peak centers at 40\,K from experiments, and are numbered at top. 
The incident energy was 100\,meV for panel (a), and 30\,meV for panel (b). 
}
\label{fig:DOS}
\end{figure*}  

\begin{figure}[t]
\includegraphics[width=0.8\columnwidth]{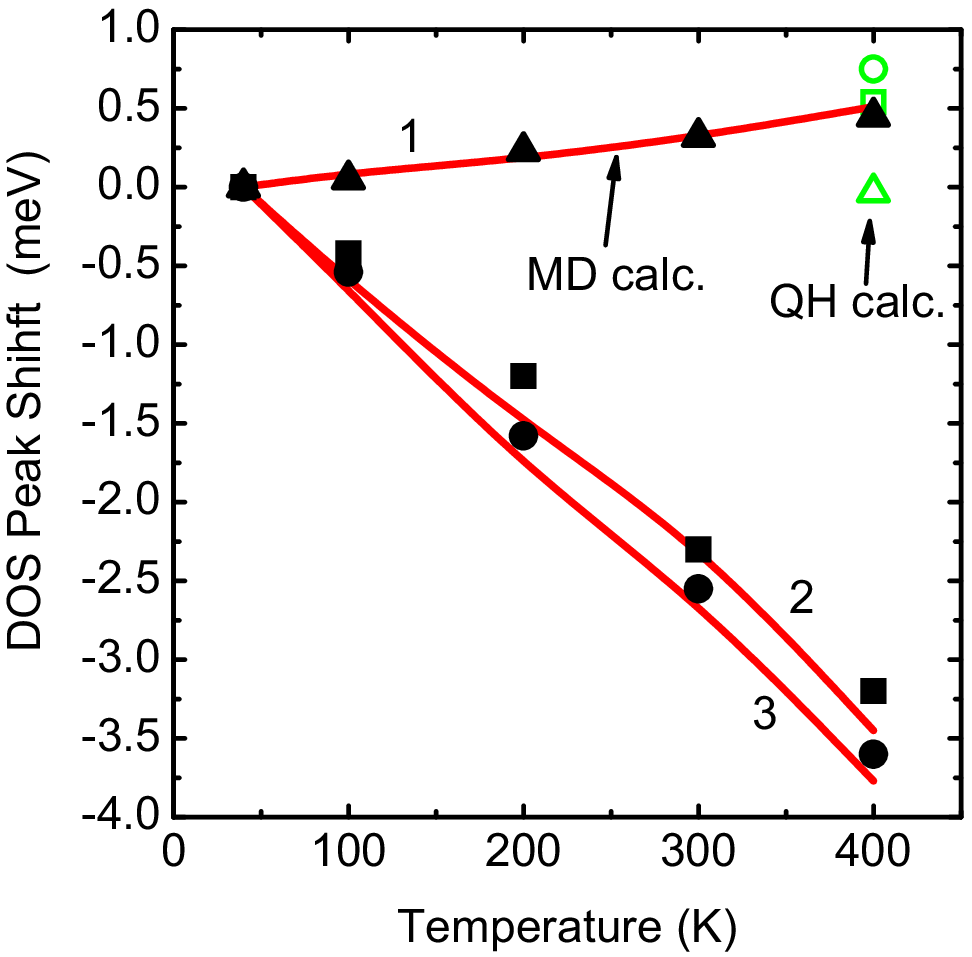}
\caption{Shifts of  centers of peaks in the phonon DOS, relative to data at 40\,K. The filled symbols are experimental data, open symbols (green) are from MD-based QHA calculations and solid curves (red) are from MD calculations. Indices 1, 2, 3 correspond to the peak labels in Fig.~\ref{fig:DOS}, and are also represented by the triangle, square and circle respectively for experimental data and QHA calcuations.}
\label{fig:DOS_fit}
\end{figure}

Figure~\ref{fig:DOS} presents the ``neutron-weighted'' phonon DOS of  Ag$_2$O with the cuprite structure from ARCS data at two incident energies at temperatures from 40 to 400\,K. 
Neutron-weighting is an artifact of inelastic neutron scattering by phonons. 
Phonon scattering scales with the scattering cross section divided by atom mass, $\sigma/m$, so the Ag-dominated modes around 8\,meV  are relatively weaker than the O-dominated modes around 63\,meV. 
Since the instrument energy resolution is inversely related to both the incident energy and the energy transfer, the spectra in Fig.~\ref{fig:DOS}(b) have generally higher resolution than in Fig.~\ref{fig:DOS}(a). 
As shown in Fig.~\ref{fig:DOS}, the main features (peaks 1, 2 and 3) of the DOS curve from inelastic neutron scattering experiments undergo substantial  broadening with temperature, even below 200\,K, indicating an unusually large anharmonicity. Along with the broadening,  peak 1 stiffens slightly, but peaks 2 and 3 shift to lower energy by more than 3.2\,meV. This is an enormous shift over such a small temperature range. Over the same temperature range, phonons of ScF$_3$ shifted by about 1 meV, for example.\cite{ScF3} To quantify thermal shifts, Gaussian functions were fitted to the three major peaks in the phonon DOS, and Fig.~\ref{fig:DOS_fit} presents the peak shifts relative to their centers at 40\,K.

\begin{figure}[]
\includegraphics[width=0.85\columnwidth]{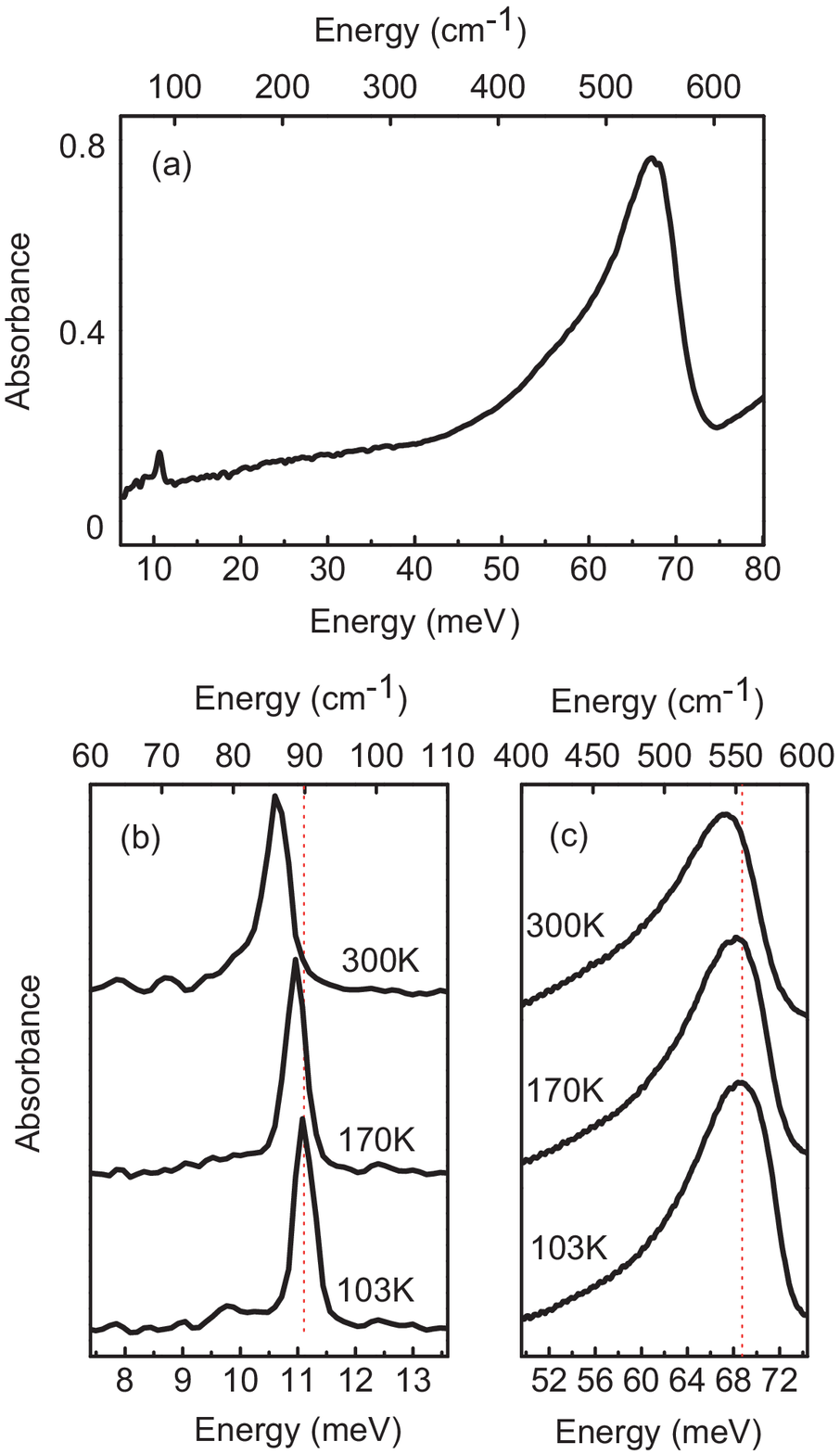}
\caption{(a) FT-IR absorbtion spectra of  Ag$_2$O with the cuprite structure at 300\,K. 
(b), (c) Enlargement of two bands at selected temperatures.}
\label{fig:FTIR}
\end{figure}

Figure~\ref{fig:FTIR} presents the infrared spectra of  Ag$_2$O  between 50 and 650\,cm$^{-1}$. Two absorption bands at 86\,cm$^{-1}$ and 540\,cm$^{-1}$ are seen, consistent with  previous measurements at  room temperature \cite{Waterhouse}. Analysis by group theory  showed  they have $F_{1u}$ symmetry. 
The low frequency mode at 86\,cm$^{-1}$ is an Ag-O-Ag bending mode, while the high frequency band corresponds to  Ag-O stretching.  
Consistent with the trend of phonon DOS measured by neutron scattering, the high frequency band  broadened significantly with temperature. From 100 to 300\,K, the two modes shifted to lower energy by about 4\,cm$^{-1}$ and 13\,cm$^{-1}$, respectively.

\section{First-Principles Molecular Dynamics Simulations}

\subsection{Methods}

First-principles calculations were performed with the generalized gradient approximation (GGA) of  density functional theory (DFT), implemented in the VASP package.\cite{vaspa,vaspb, vaspc}
Projector augmented wave pseudopotentials and a plane wave basis set with an energy cutoff of 500 eV were used in all calculations.

First-principles Born-Oppenheimer molecular dynamics simulations were performed for a $3 \times 3 \times 3$ supercell with temperature  control by a Nos{\'{e}} thermostat. The relatively small simulation cell could be a cause for concern,\cite{sizeeffect} 
but  convergence testing showed that the supercell in our study is large enough to accurately capture the phonon anharmonicity of Ag$_2$O. The simulated temperatures included 40, 100, 200, 300 and 400\,K. 
For each temperature, the system was first equilibrated for 3 ps, then simulated for 18 ps with a time step of 3\,fs. The system was fully relaxed at each temperature, with convergence of the pressure within 1 kbar.

Phonon frequency spectra and their $k$-space structure were obtained from the MD trajectories
by the Fourier transform velocity autocorrelation method. \cite{autocorra,autocorrb, autocorrc, autocorrd}
The phonon DOS is  
\begin{equation}
\label{eq:DOScal}
g(\omegaup)=\sum_{n,b} \,\int dt\, e^{-{\rm i} \omegaup t}  \langle \vec{v}_{n,b}(t)\,\vec{v}_{0,0}(0) \rangle
\end{equation}
where $\langle \, \rangle$ is an ensemble average, and $\vec{v}_{n,b}(t)$ is the velocity of the atom $b$ in the unit cell $n$ at time $t$.
Individual phonon modes could also be projected
onto each $k$-point in the Brillouin zone by computing the phonon power spectrum. \cite{autocorrb, autocorrd} 
To better compare  with  data from inelastic neutron scattering, the calculated DOS at each temperature was convoluted with the ARCS instrumental broadening function, and was neutron-weighted appropriately. \cite{ScF3, datareduce} 

Calculations in the quasiharmonic approximation were performed  two ways. 
In the lattice dynamics method, the thermal expansion was evaluated by optimizing the 
vibrational free energy as a function of volume. \cite{ScF3}
Calculations were performed self-consistently with a 6-atom unit cell with a $10\times 10 \times 10$ $k$-point grid. 
Phonon frequencies were calculated using the small displacement method implemented by the Phonopy package.\cite{phonopy}
These  phonon dispersions in the QHA were also used for the anharmonic perturbation theory described below. 
The second method used MD calculations to implement the QHA. 
We  removed the temperature-dependent explicit anharmonicity by 
performing simulations at 40\,K for volumes characteristic of 400\,K, 
which produced a pressure of 0.45\,GPa at 40\,K. 
Further computational details are given in the Supplemental Material. \footnote{See Supplemental Material at [URL will be inserted by publisher] for the description of the MD-implemented QHA and the corresponding figure.}

At the lowest simulation temperature of 40 K, classical MD trajectories may require justification. In principle,  nuclear motions could be better treated by mapping each nucleus onto a classical system of several fictitious particles governed by an effective Hamiltonian, derived from a Feynman path integral, for example. \cite{quantum1, quantum2} Such low temperature quantum effects are beyond the scope of this work. Nevertheless, our results should not be altered significantly by quantum effects for the following reasons. Our particular interest is in anharmonic phonon-phonon interactions at higher temperatures, and our new results concern the phonons and NTE above 250 K. A classical MD simulation is usually appropriate at higher temperatures. The modes most subject to quantum corrections are those involving the dynamics of the lower mass O atoms, but these are at high energies. They are not activated at 40 K, and show weak anharmonic effects. Relatively larger anharmonic effects at low temperatures are found in the modes below 10 meV. These are dominated by the Ag atoms, but with their high mass only tiny quantum effects are expected. There are several semi-quantitative methods to estimate the magnitudes of  quantum corrections. For example, Berens, \textit{et al}. \cite{quantum3} and Lin, \textit{et al}. \cite{quantum4} suggest that quantum effects could be evaluated from the difference between the quantum and classical vibrational energy or free energy derived from the corresponding partition functions. These methods do not account for all quantum effects on nuclear trajectories, but for Ag$_2$O at 40 K, by using both classical and quantum paritition function for the same phonon DOS, we found the vibrational energy difference between classical and quantum statistics is only 1.2\% of the cohesive energy.

\subsection{Results}

\begin{figure}[b]
\includegraphics[width= 0.8\columnwidth]{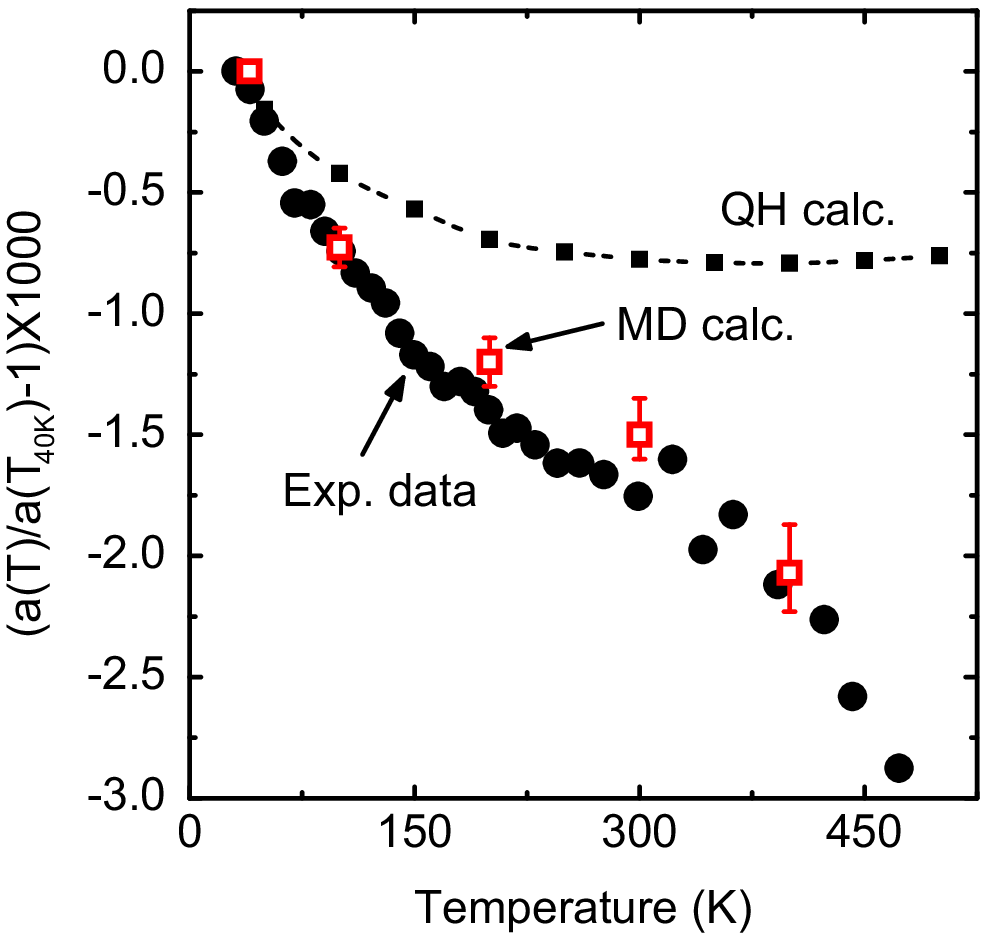}
\caption{Temperature dependence of  lattice parameter from experimental data in Ref. [\onlinecite{NTEa}], quasiharmonic calculations and MD calculations, expressed as the relative changes with respect to their 40\,K values, i.e., $a(T)/a(40\,K)-1$. }
\label{fig:expansion}
\end{figure}  


\begin{table}[t]
\caption{\label{tab:table1} Properties of  Ag$_2$O with the cuprite structure from present MD calculations, compared to experimental data. 
Units: lattice parameters in \AA, bulk modulus in GPa, 
thermal expansion coefficients in 10$^{-6}$K$^{-1}$, 
vibrational frequencies in meV.
}
\begin{ruledtabular}
\begin{tabular}{ccc}
& Experiment \footnote{Lattice parameter at 40\,K is from neutron scattering measurements in the present work, which is in good agreement with Refs. [\onlinecite{NTEa, NTEb}]. 
Bond linear thermal expansion (LTE) and variance data are from Refs. [\onlinecite{exafsb, chapman}]. 
The IR active mode frequencies are from FT-IR measurements in the present work, and the frequencies of the lowest two modes are from the luminescence spectra in Ref. 
[\onlinecite{lumine}].} 
& Calculation\\\hline
Lattice Parameter&&\\
\!\!\!a  &4.746 &4.814\\

Bulk Modulus&&  \\
\!\!K & N/A  &72      \\

Bond LTE&&\\
 \!\!$\betaup_{Ag-O}$  & 12.1--35.0&19.4\\
 $\betaup_{Ag-Ag}$& -9.99&-14.7\\
Bond Variance&&\\
$\sigma_{Ag-Ag}$ (40\,K)& 0.0078&0.0073\\
\,\,\,$\sigma_{Ag-Ag}$ (400\,K)& 0.053&0.067\\
Mode Frequency&&\\
$F_{2u}$ &5.60 &5.58 \\
$E_u$ &8.9 &7.45\\
$F_{1u}^{(1)}$ & 11.2 & 11.5\\
$A_{2u}$ & N/A& 29.4\\
$F_{2g}$ &N/A &48.4 \\
$F_{1u}^{(2)}$ & 67 & 63.6\\
 
\end{tabular}
\end{ruledtabular}
\end{table}

Table~\ref{tab:table1} presents results from our MD simulations and experimental data on lattice parameter, bulk modulus, thermal expansivity, and phonon frequences at the $\Gamma$-point. 
As shown in Fig.~\ref{fig:expansion}, the MD simulations predicted the NTE  very well. 
On the other hand, consistent with a recent QHA calculation, \cite{inelasticneutron} 
the NTE calculated with the QHA 
method was much smaller. 

 

The phonon DOS curves calculated from first-principles MD simulations 
are shown in Fig.~\ref{fig:DOS}
with the experimental spectra for comparison. 
To facilitate visual comparison, in Fig.~\ref{fig:DOS}(a) the energy axis of the calculated spectra were scaled by 6.8\% to correct for underestimates of the force constants in the GGA method.
Nevertheless, excellent agreement is found between the simulated phonon DOS and the experimental data,
and the calculated thermal broadenings and shifts are in good agreement, too. 
Gaussian functions were also fit to the calculated spectra,
and Fig.~\ref{fig:DOS_fit} compares these thermal shifts from  experiment and  calculation.

Because of the large mass difference between Ag and O atoms, 
the O-dominated phonon modes are well separated from the Ag-dominated modes. 
Partial phonon DOS analysis showed that
the Ag-dominated modes had similar energies, forming the peak of the phonon DOS below 20\,meV (peak 1 in Fig.~\ref{fig:DOS}), whereas the O-dominated modes had energies above 40\,meV (peaks 2 and 3).

\begin{figure}[t]
\includegraphics[width=0.8\columnwidth]{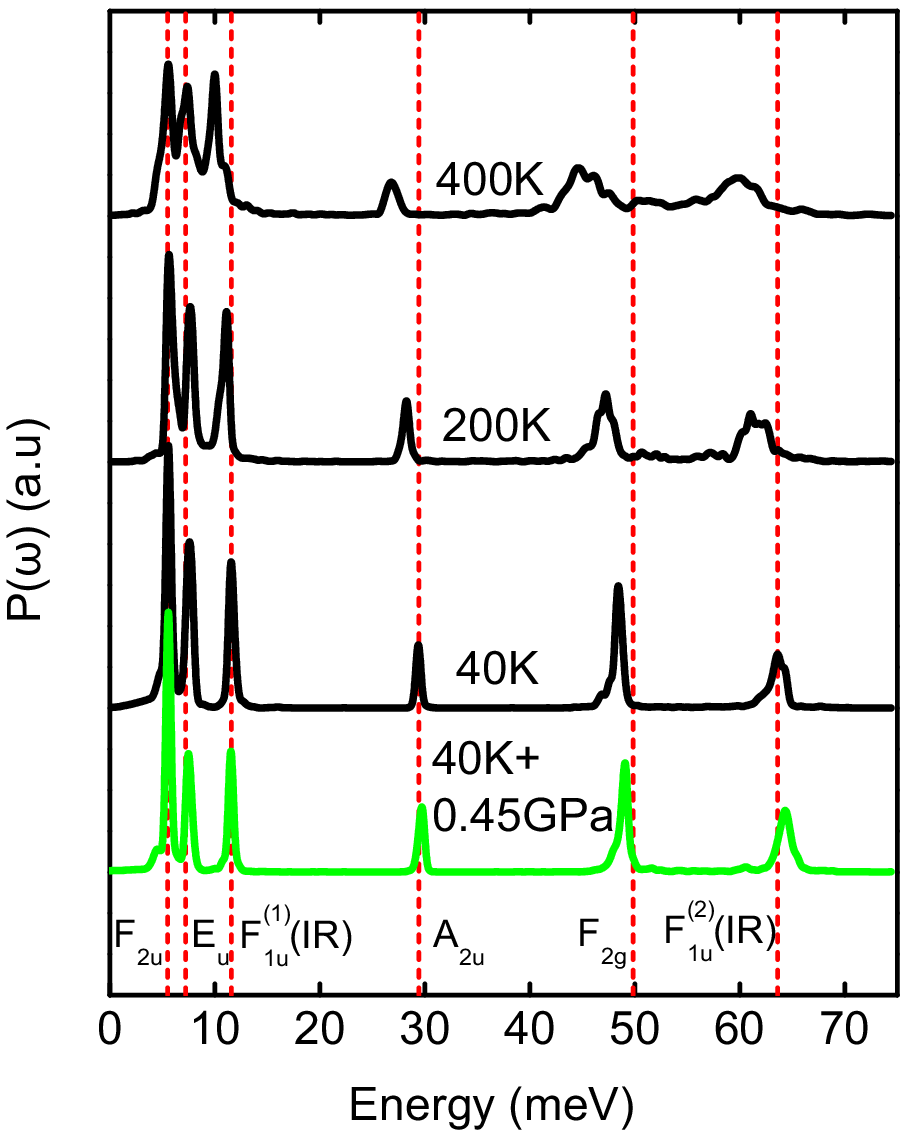}
\caption{Phonon modes simulated
by MD and projected on the $\Gamma$-point, at temperatures and pressures as labeled. The normal-mode frequencies calculated from harmonic lattice dynamics are shown as vertical dashed lines in red. The group symmetry for each mode is shown at the bottom.}
\label{fig:PSD}
\end{figure}

\begin{figure}[]
\includegraphics[width=0.8\columnwidth]{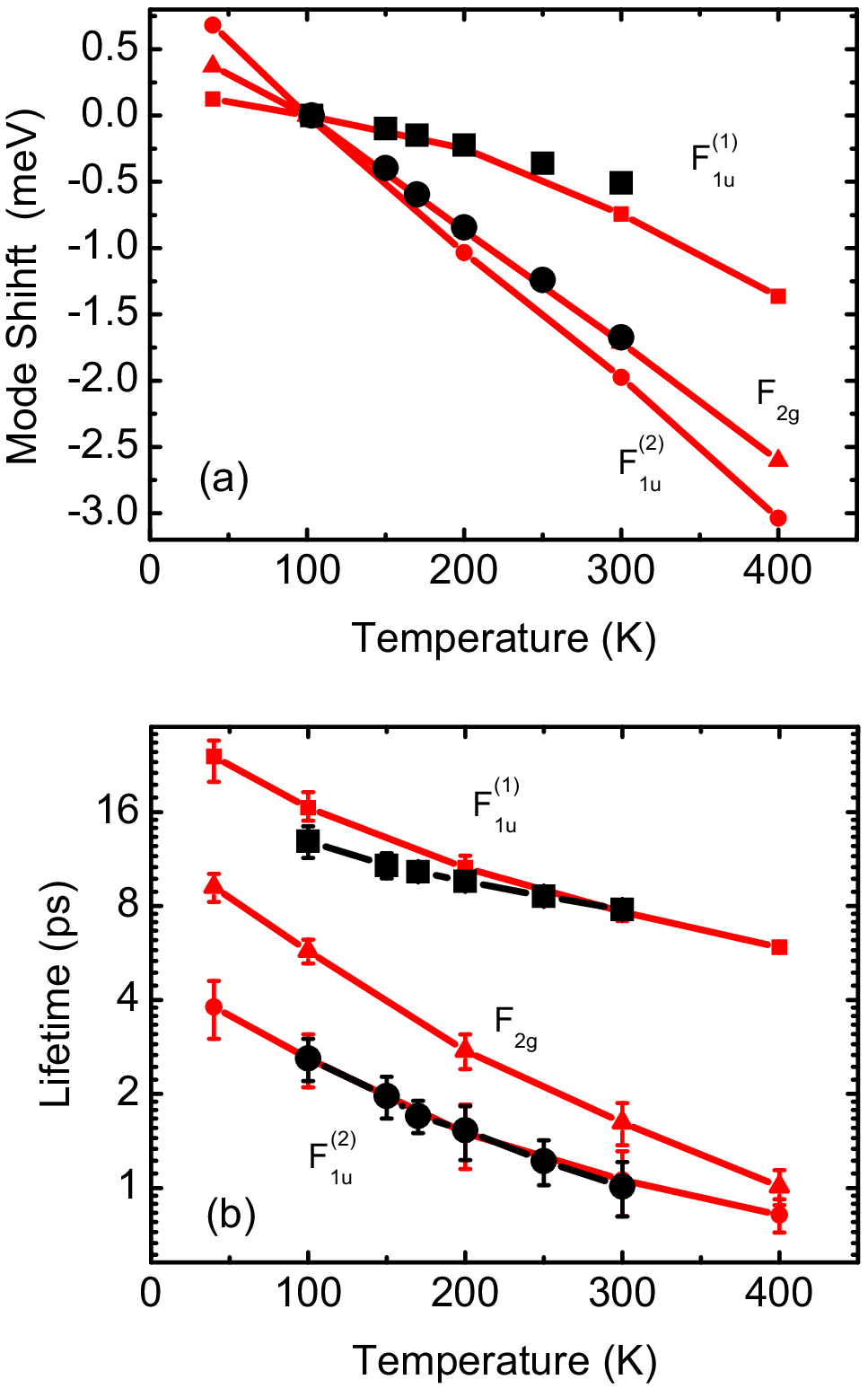}
\caption{(a) Temperature dependent frequency shifts of the Ag-dominated  $F_{1u}^{(1)}$ mode and the O-dominated  $F_{2g}$ and $F_{1u}^{(2)}$ modes from FT-IR (black), compared with the MD simulated peaks (red) such as in Fig.~\ref{fig:PSD}. 
(b) The lifetimes of the corresponding modes at temperatures from 40 to 400 \,K, from FT-IR (black) and the MD simulated peaks (red).}
\label{fig:PSD_fit}
\end{figure}

Figure~\ref{fig:PSD} shows the normal modes from MD simulations, projected to the $\Gamma$-point. 
Six vibrational modes are evident, including the IR-active  $F_{1u}^{(1)}$ and $F_{1u}^{(2)}$ modes. 
The calculated frequencies of these modes at 40\,K are listed in Table~\ref{tab:table1}, showing good agreement with experiment. Large thermal shifts and broadenings are apparent in the 
simulated frequencies, consistent with experiment. 
The calculated peaks were then fitted with Lorentzian functions to extract the centroids and linewidths, 
which compare well with the
FT-IR data  as shown in Fig.~\ref{fig:PSD_fit}.  

\section{Anharmonic Perturbation Theory}
 
\subsection{Computational Methodology}

Cubic anharmonicity gives rise to three-phonon processes,
which are an important  mechanism of phonon-phonon interactions.
The strengths of 
the three-phonon processes depend on two elements -- 
the cubic anharmonicity tensor that gives the coupling strengths between three phonons,
and the kinematical processes described by the two-phonon density of states (TDOS). \cite{cubic1, cubic2, cubic3} 

From Ipatova, \textit{et al.} \cite{tensor1}, an anharmonic tensor element 
for a process with the initial phonon mode $j$ at the $\Gamma$-point and $s$ phonons is
\begin{eqnarray}
\label{eq:tensor}
V(j;{\vec{q}_1}j_1;...;{\vec{q}_{s-1}}j_{s-1})&=&\frac{1}{2s!}\left(\frac{\hbar}{2N}\right)^{\frac{s}{2}}N \, \Delta({\vec{q}}+{\vec{q}_1}+\cdots+{\vec{q}_{s-1}})\nonumber \\
\times[\omegaup\,\omegaup_1 \cdots &\omegaup_{s-1}&]^{\frac{1}{2}}C(j;{\vec{q}_1}j_1;...;{\vec{q}_{s-1}}j_{s-1})
\end{eqnarray} 
where the phonon modes $\{ \vec{q}_i j_i \}$ 
have quasiharmonic frequencies $\{ \omegaup_i \}$ and occupancies $\{ n_i \}$. 
The $C(.)$'s are expected to be slowly-varying functions of their arguments. \cite{tensor2} 
We assume the term $C(j;{\vec{q}_1}j_1;...;{\vec{q}_{s-1}}j_{s-1})$ 
is a constant for the initial phonon $j$, and use it as a  parameter when fitting to  trends from MD or experiment. 
Although  
$C(j;{\vec{q}_1}j_1;{\vec{q}_{2}}j_{2})$ 
changes with
$j_1$ and $j_2$, an average over modes, 
\begin{eqnarray}
\langle C(.) \rangle = \frac{\sum_{1,2} C(j;{\vec{q}_1}j_1;{\vec{q}_{2}}j_2)}{\sum_{1,2} 1}
\end{eqnarray}
is found by fitting to experimental or simulational results, where 1, 2 under the summation symbol represent ${\vec{q}_i}j_i$.
We define the cubic fitting parameter as
\begin{equation}
\label{eq:fitpara}
C^{(3)}_j = \langle C(j;{\vec{q}_1}j_1;{\vec{q}_2}j_2) \rangle 
\end{equation}

The second key element  of perturbation theory is that interacting phonons satisfy the kinematical conditions of 
conservation of energy and momentum.
This condition is averaged over all phonons with 
the two-phonon density of states (TDOS), defined as
\begin{eqnarray}
\label{eq:D}
&&D(\omegaup)=\sum_{{\vec{q}_1},j_1} \sum_{{\vec{q}_2}j_2}D(\omegaup,\omegaup_1,\omegaup_2)  \nonumber \\
&=&\frac{1}{N}\sum_{{\vec{q}_1},j_1} \sum_{{\vec{q}_2},j_2}\Delta({\vec{q}_1}+{\vec{q}_2}) \, 
 \big[(n_1+n_2+1) \, \delta(\omegaup-\omegaup_1-\omegaup_2)\nonumber \\
&\mbox{}& \quad \quad \quad \quad \quad + \, 2 (n_1-n_2) \, \delta(\omegaup+\omegaup_1-\omegaup_2) \big]
\end{eqnarray} 
The first and second terms in square brackets
are from down-conversion  and up-conversion scattering processes, respectively. 

The strength of the cubic phonon anharmonicity can be quantified by the quality factor ${\mathcal Q}$, related to the phonon lifetime as the number of the vibrational periods for the energy to decay to a factor of $1/$e,
and ${\mathcal Q} = \omegaup / 2 \Gamma$, where $2 \Gamma$ is the linewidth of the phonon peak. 
Considering Eqs.~(\ref{eq:tensor}) to (\ref{eq:D}), the phonon linewidth is related to the TDOS, $D(\omegaup)$, weighted by the coupling strength.\cite{{tensor2,autocorrd}} To leading order, the inverse of the quality factor can be expressed as a function of the TDOS
\begin{equation}
\label{eq:fitting}
\frac{1}{{\mathcal Q}_j} = \frac{\pi\hbar}{32} |C^{(3)}_j|^2 \sum_{{\vec{q}_1},j_1} \sum_{{\vec{q}_2},j_2}\omegaup_1\omegaup_2  \, D(\omegaup,\omegaup_1,\omegaup_2) 
\end{equation}
The TDOS at various temperatures was calculated from the kinematics of all three-phonon processes, sampling the phonon dispersions with a $16\times 16 \times 16$ $q$-point grid for good convergence. 
The  ${\mathcal Q}$ from MD simulations were used to approximate the anharmonicity of the phonon modes
of different energies, 
and obtain the the coupling strengths $|C^{(3)}_j|^2$ for the different modes. 

\subsection{Results}

\begin{figure}[t]
\includegraphics[width=0.8\columnwidth]{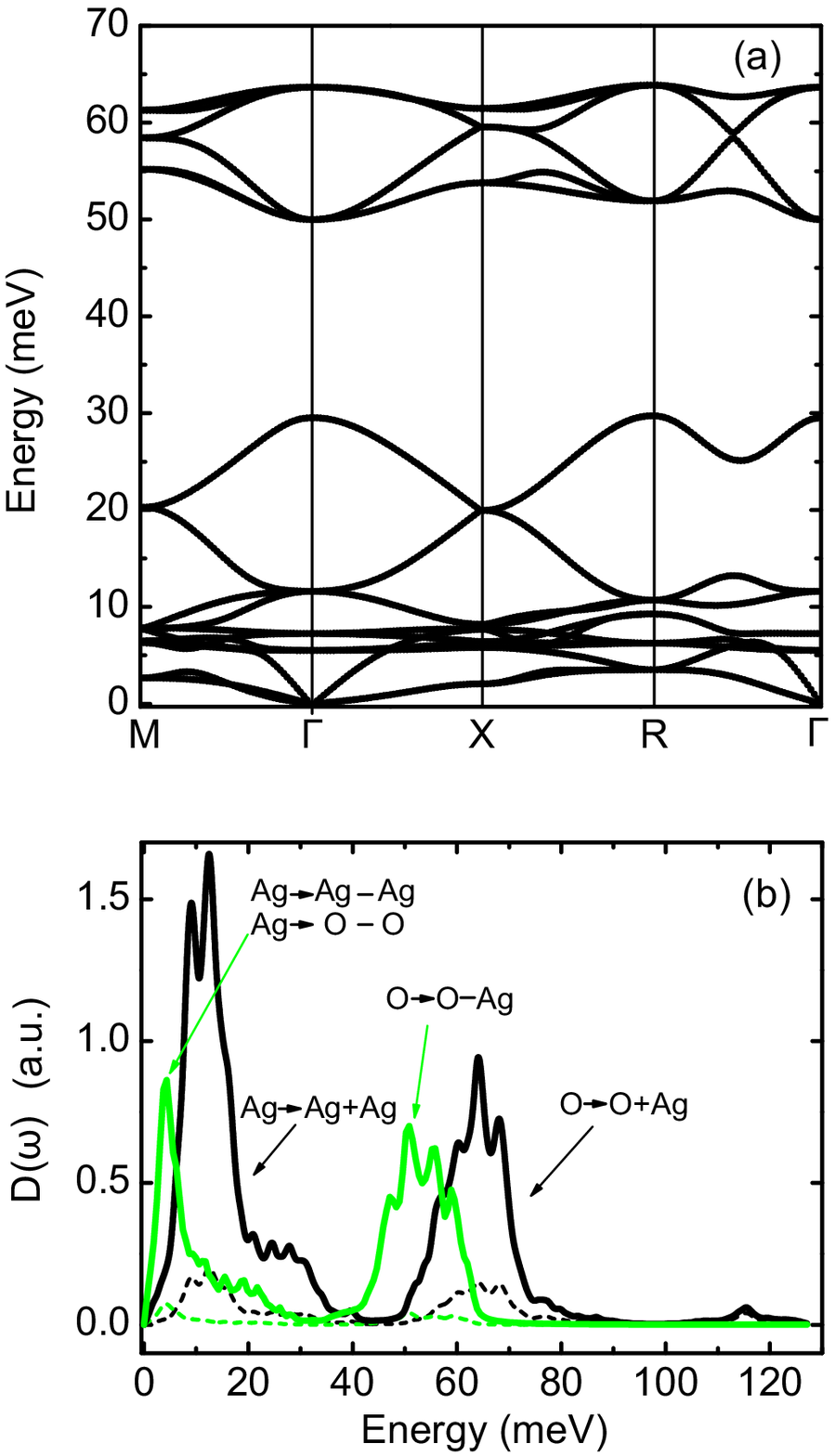}
\caption{(a) Calculated phonon dispersion along
high-symmetry directions of  Ag$_2$O with the cuprite structure. $\Gamma\, (0,0,0)$, $M \,(0.5,0.5,0)$, $X \, (0.5,0,0)$, $R \,(0.5,0.5,0.5)$.
(b) The TDOS spectra, $D(\omegaup)$, at 40\,K (dashed) and 400\,K (solid). The down-conversion and up-conversion contributions are  presented separately as  black and green curves, respectively. }
\label{fig:2DOS}
\end{figure} 

\begin{figure}[t]
\includegraphics[width=1\columnwidth]{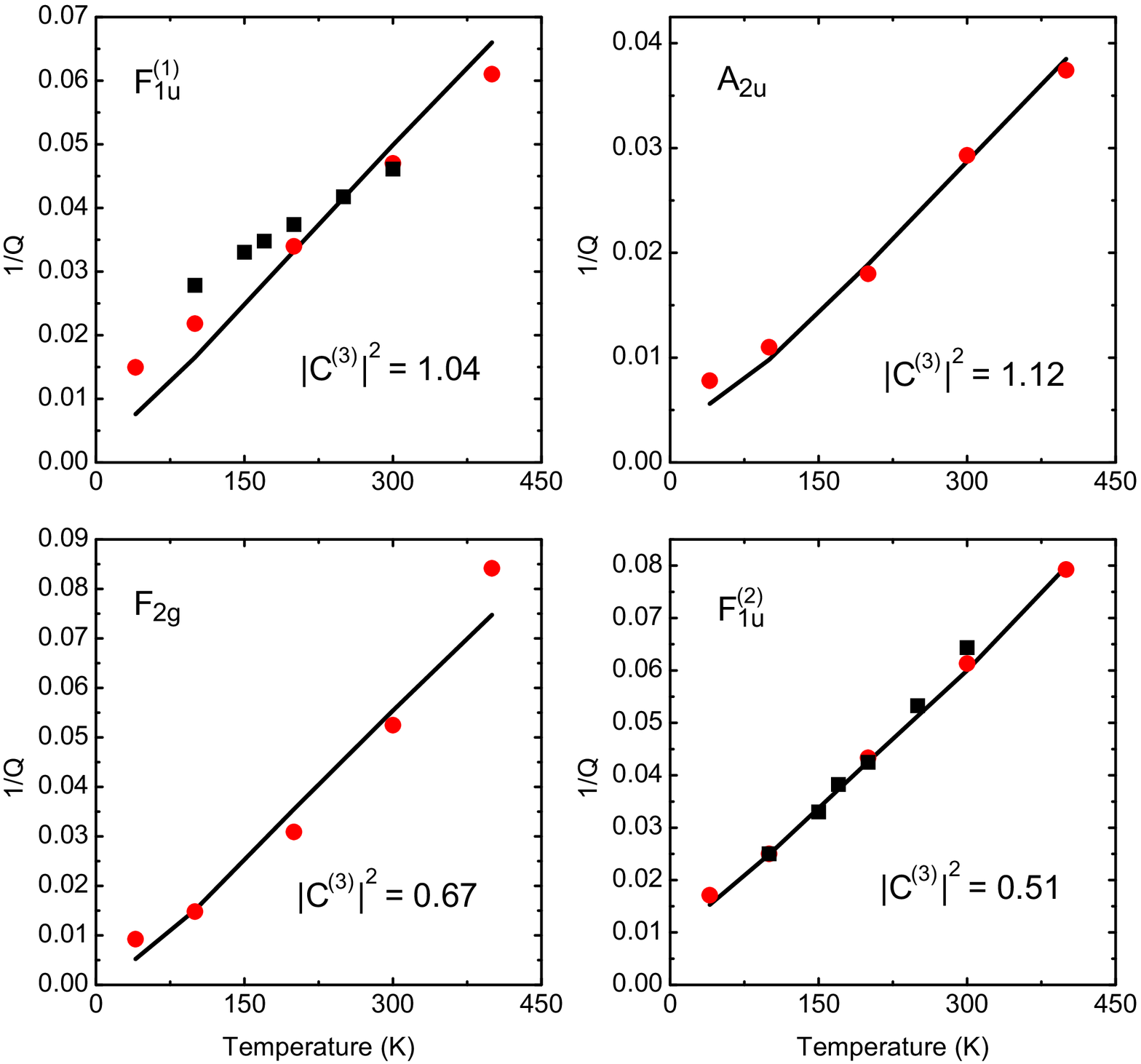}
\caption{Temperature dependence of the inverse of the quality factors $1/{\mathcal Q}$ for the $F_{1u}^{(1)}$, $A_{2u}$, $F_{2g}$ and $F_{1u}^{(2)}$ phonon modes from FT-IR (black squares), MD calculations(red circles) and the theoretical fittings with a full calculation of the TDOS. The unit of the fitting parameter $|C_j^{(3)}|^2$ is 10$^{-1}$ eV$^{-1}$. }
\label{fig:mode_fit}
\end{figure} 

Fig.~\ref{fig:2DOS}(a) shows  calculated phonon dispersion curves of  Ag$_2$O with the cuprite structure along high-symmetry directions.
From these,  the TDOS spectra, $D(\omegaup)$, were obtained at different temperatures, presented in Fig.~\ref{fig:2DOS}(b) for 40 and 400\,K. 
At low temperatures there are two small peaks in the TDOS centered at 15 and 65\,meV. 
Our calculation showed that the peak at 15\,meV is from the decay processes of one Ag-dominated mode into two with lower frequencies, i.e., Ag $\mapsto$ Ag $+$ Ag. The peak at 65\,meV originates from spontaneous decay of one O-dominated mode into another O-dominated mode of lower frequency and one Ag-dominated mode.

At high temperatures there are more down-conversion processes, but an even greater change in up-conversion processes. 
Figure~\ref{fig:2DOS}(b) shows how the strong down-conversion peaks at low temperatures grow approximately linearly with temperature, following the thermal population of phonon modes involved in the interactions. Near the peak at 65\,meV, one up-conversion band centered at 50\,meV is also strong. This band comprises scattering channels in which one O-dominated mode is combined with a Ag-dominated mode to form a higher frequency O-dominated mode, i.e., O $\mapsto$ O $-$ Ag. 
At the low energy side, there is another band below 15\,meV from two types of  up-conversion processes. One is from Ag-dominated modes alone, i.e., Ag $\mapsto$ Ag $-$ Ag.
The other involves two O-dominated modes, i.e., Ag $\mapsto$ O $-$ O, owing to the increased number of higher energy O-dominated modes that can participate in these processes at higher temperatures. 

Figure~\ref{fig:mode_fit} shows the inverse of quality factors, $1/{\mathcal Q}$, of the $F_{1u}^{(1)}$, $F_{2g}$, $F_{1u}^{(2)}$ and  $A_{2u}$ phonon modes from FT-IR and MD calculations, together with the best theoretical fits with Eq.~(\ref{eq:fitting}). These modes are near the centers of main features of the phonon DOS shown in Fig.~\ref{fig:DOS}, i.e., the peaks 1, 2, 3 and the gap in between, respectively, and are useful for understanding the overall anharmonicity. As shown in Fig.~\ref{fig:mode_fit}, at higher temperatures the quality factors decrease substantially. At 400\,K, the  $F_{1u}^{(1)}$, $F_{2g}$, $F_{1u}^{(2)}$ modes have low ${\mathcal Q}$ values from 12 to 15, but the  $A_{2u}$ mode has a much larger value of  26. With a single parameter $|C_j^{(3)}|^2$ for each mode, good fittings to the quality factors are obtained. The fitting curves and the corresponding values of the parameters $|C_j^{(3)}|^2$ are presented in Fig.~\ref{fig:mode_fit}. The $|C_j^{(3)}|^2$ values do not vary much among different modes, so the typical assumption of a slowly varying  C(.) seems reasonable. 


\section{Discussion}

\subsection{Quasiharmonic Approximation}

In the quasiharmonic approximation (QHA),
a mode Gr\"{u}neisen parameter $\gammaup_j$ is defined as the ratio of the fractional  
change of the mode frequency $\omegaup_j$ to the fractional change of volume $V$, at constant temperature,
$\gammaup_j = - \frac{\partial(\ln \omegaup_j)}{\partial (\ln V)}$.
The usual trend is for phonons to soften with lattice expansion, increasing the phonon entropy and
stabilizing the expanded lattice at elevated temperatures.
A negative Gr\"{u}neisen parameter is therefore expected
for the special phonon modes associated with NTE, such as RUMs. 
If all the anharmonicity of  Ag$_2$O with the cuprite structure is attributed to this volume effect, however,  
the values of Gr\"{u}neisen parameters are approximately --9 for the high energy modes at peaks 2 and 3 of the phonon DOS, and --20 for the infrared-active F$_{1u}$ mode. These anomalous values may indicate a problem with the QHA. The QHA method also significantly underestimates the NTE, and misses the behavior at temperatures above 250\,K.



As seen in the figure in the Supplemental Material and in Fig.~\ref{fig:PSD}, 
the giant negative Gr\"{u}neisen parameters are inconsistent with the
results of MD simulations. 
The volume change 
alone does not affect much the phonon lifetimes or frequency shifts.  
All features in the phonon spectra generated with the MD-implemented QHA
showed little change except for small stiffening about 0.6\,meV at high energies, 
in agreement with the recent lattice dynamics QHA calculations. \cite{inelasticneutron}

\subsection{Negative Thermal Expansion}


Our eigenvector analysis of phonon modes showed that the three low-energy Ag-dominated $F_{2u}$, $E_u$ and $F_{1u}$ modes correspond to two distinct types of vibrations. The  $F_{2u}$ mode involves  rigid rotations of Ag$_4$O tetrahedra, and could be considered as RUM. The  $E_u$ mode involves the shearing of Ag$_4$O units by changing the Ag-O-Ag bond angles. The  $F_{1u}$ mode measured by infrared spectrometry also shears the Ag$_4$O units, and includes some displacements of O atoms.  Shearing the tetrahedra was shown to reduce the average vertex-vertex distance \cite{Waterhouse,tension} and contribute to the NTE. 

In  the cuprite structure, the modes associated with the rigid rotations and the distortions of the Ag$_4$O tetrahedra have similar energies below 10\,meV.  They are equally favorable thermodynamically, and both would be active at very low temperature. As a consequence, there is simultaneously a large deformation of Ag$_4$O units and a strong contraction of the Ag-Ag shell, as observed experimentally and computationally at low temperatures. 
The large thermal distortions of the Ag$_4$O tetrahedra  involve the Ag atoms at the vertices, 
and their large mass causes these distortions to occur at low frequencies. Polyhedral units in most other NTE materials are bridged by the lightweight atoms, such as O and C-N, so the polyhedra are distorted at significantly higher energies (usually above 40\,meV), and may not distort at lower temperatures. 

It is a thermodynamic requirement that the NTE of Fig. \ref{fig:expansion} must go to zero at $T=0$, but the intervening phase transition at 40\,K impedes this measurement. Nevertheless, the steep slope of the lattice parameter with temperature is consistent with the occupancy of phonon modes of 10\,meV energy, suggesting that the QHA model of NTE involves the correct modes,
such as the $F_{2u}$ mode (which is a RUM). These low-energy modes are dominated by motions of the Ag atoms. 
This explanation based on the QHA is qualitatively correct, but 
anharmonic interactions are large enough to cause the QHA to underestimate
the NTE by a factor of two.

\subsection{Explicit Anharmonicity}

At temperatures above 250\,K, there is a second part of the 
NTE behavior that is  beyond the predictions of  quasiharmonic theory.
This NTE above 250\,K is predicted accurately by the ab-initio MD calculations,
so it is evidently a consequence of phonon anharmonicity. 
The temperature-dependence of this NTE behavior follows the Planck 
occupancy factor for phonon modes above 50\,meV, corresponding to the O-dominated band of optical frequencies.
In the QHA these modes above 50\,meV do not contribute to the NTE.
These modes are highly anharmonic, however, as shown by their large broadenings and shifts. 
 
For  cubic anharmonicity, the two-phonon DOS (TDOS) is the spectral quantity parameterizing the 
number of phonon-phonon interaction channels available to a phonon. 
For  Ag$_2$O with the cuprite structure, the peaks in the TDOS overlap well with the peaks in the phonon DOS. 
Most of the phonons therefore have many possible interactions with other phonons, which contributes to the large anharmonicity of 
 Ag$_2$O with the cuprite structure, and small ${\mathcal Q}$  (short lifetimes). 
Although the ${\mathcal Q}$ values of most phonon modes in  Ag$_2$O with the cuprite structure
are small and similar, the origins of these lifetime broadenings are intrinsically different. 
For  peak 2 of the phonon DOS, the anharmonicity is largely from the up-conversion 
processes: O $\mapsto$ O $-$ Ag, while for peak 3 it is from the down-conversion 
processes: O $\mapsto$ O $+$ Ag. 
The anharmonicity of peak 1 is more complicated. 
It involves both  up-conversion and down-conversion processes of Ag-dominated modes. 
The TDOS also shows why the  $A_{2u}$ mode has a larger ${\mathcal Q}$  than other modes. 
Figure~\ref{fig:2DOS}(b) shows that the  $A_{2u}$ mode lies in the trough of the TDOS 
where there are only a few  phonon decay channels. 

Owing to explicit anharmonicity from phonon-phonon interactions, the thermodynamic 
properties of  Ag$_2$O with the cuprite structure cannot be understood as a sum of contributions from independent normal modes. 
The frequency of an anharmonic phonon depends on the level of excitation of other modes. 
At high temperatures, large vibrational amplitudes increase the anharmonic coupling of modes, 
and this increases the correlations between the motions of the Ag and O atoms, as shown by perturbation theory. 
Couplings in perturbation theory have phase coherence, 
so the coupling between Ag- and O-dominated modes at higher energies,
as seen in the peak of the TDOS,
causes correlations
between the motions of Ag and O atoms. 
The ab-initio MD simulations
show that anharmonic interactions allow 
the structure to become more compact with increasing vibrational amplitude. 
The mutual motions of the O and Ag atoms cause higher density as the atoms fill space more effectively.
The large difference in atomic radii of Ag and O may contribute to this effect.
Perhaps it also facilitates the irreversible changes in Ag$_2$O
at temperatures above 500\,K, but this requires further investigation. 
For cuprite Cu$_2$O, which has less of a difference in atomic radii, 
the thermal expansion is much less anomalous.

\section{Conclusions}
Phonon densities of states of  Ag$_2$O with the cuprite structure at temperatures from 40 to 400\,K were measured by inelastic neutron scattering spectrometry. The infrared spectra of phonon modes were also obtained at temperatures from 100 to 300\,K. Large anharmonicity was found from both the shifts and broadenings of peaks in the phonon spectra.
A normal mode analysis identified the rigid unit modes and the bending modes of the Ag$_4$O tetrahedra that play key roles in the negative thermal expansion (NTE) at low temperatures. 
Some of the NTE can be understood by quasiharmonic theory, but this approach 
is semiquantitative, and limited to temperatures below 250\,K.

First principles MD calculations were performed at several temperatures. These calculations accurately accounted for the NTE and local dynamics of  Ag$_2$O with the cuprite structure, such as the contraction of the Ag-Ag shell and the large distortion of the Ag$_4$O tetrahedra. 
The phonon DOS obtained from a Fourier-transformed velocity autocorrelation method showed  large anharmonic effects in Ag$_2$O, in excellent agreement with the experimental data. 
A second part of the NTE at temperatures above 250\,K is due largely to the anharmonicity of phonon-phonon interactions 
and is not predicted with volume dependent quasiharmonicity. 

Phonon perturbation theory with the cubic anharmonicity
helped explain the effects of phonon kinematics on phonon anharmonicity of  Ag$_2$O with the cuprite structure. 
The phonon interaction channels for three-phonon processes are given by the TDOS, weighted approximately by the phonon coupling strength. The phonons that are most broadened are those with energies that lie on peaks in the TDOS.
The temperature-dependence of the quality factors ${\mathcal Q}$ of individual phonon modes measured by infrared spectrometry were explained well by anharmonic perturbation theory.
Perturbation theory also showed strong interactions between the Ag-dominated modes and the O-dominated modes in both up-conversion and down-conversion processes. 
In particular, the strong interactions of O-dominated modes with Ag-dominated modes 
causes the second stage of NTE at temperatures above 250\,K.

\begin{acknowledgments}
This work was supported by DOE BES under contract DE-FG02-03ER46055. The work benefited from software developed in the DANSE project under NSF award DMR-0520547. Research at Oak Ridge National Laboratory's SNS was sponsored by the Scientific User Facilities Division, BES, DOE.
\end{acknowledgments}

\bibliography{ref}

\clearpage
 
\end{document}




\title{Anharmonic lattice dynamics of Ag$_2$O studied by inelastic neutron scattering and first principles molecular dynamics simulations\\
Supplemental Material}



\author{Tian Lan} 
\email[]{tianlan@caltech.edu}
\affiliation{\mbox{Department of Applied Physics and Materials Science, \nolinebreak California Institute of Technology, Pasadena, California 91125, USA}}
\author{Chen W. Li}
\affiliation{Oak Ridge National Laboratory, Oak Ridge, Tennessee 37831, USA}
\author{J. L. Niedziela}
\affiliation{Oak Ridge National Laboratory, Oak Ridge, Tennessee 37831, USA}
\author{Hillary Smith}
\affiliation{\mbox{Department of Applied Physics and Materials Science, \nolinebreak California Institute of Technology, Pasadena, California 91125, USA}}
\author{Douglas L. Abernathy}
\affiliation{Oak Ridge National Laboratory, Oak Ridge, Tennessee 37831, USA}
\author{George R. Rossman}
\affiliation{Division of Geological and Planetary Sciences, California Institute of Technology, Pasadena, California 91125, USA} 
\author{Brent Fultz}
\affiliation{\mbox{Department of Applied Physics and Materials Science, \nolinebreak California Institute of Technology, Pasadena, California 91125, USA}}

\date{\today}




\maketitle


To separate the effects of quasiharmonicity and explicit anharmonicity,
the mode frequency $\omegaup_j=\omegaup_j(V,T)$ is expressed as a function 
of volume and temperature
\begin{equation}
\label{eq:weight}
\left(\frac{\partial \ln \omegaup_j}{\partial T}\right)_P = - \gammaup_j\betaup +
\left(\frac{\partial \ln \omegaup_j}{\partial T}\right)_V
\end{equation}
where $j$ is the phonon mode index, $\betaup$ is the volume thermal expansivity 
and $\gammaup_j$ is the mode Gr\"{u}neisen parameter. 
The left-hand side 
is the temperature-dependent  frequency shift at constant pressure, 
and includes contributions from both  quasiharmonicity and explicit anharmonicity. 
The first term on the right-hand side, the isothermal frequency shift as a function of pressure, 
is the quasiharmonic contribution to the frequency shift. 
The second term on the right is the pure temperature 
contribution to the frequency shift from the explicit anharmonicity.
From the difference of the isobaric and isothermal frequency shifts, 
the explicit anharmonicity can be determined.

\begin{figure}[b]
\includegraphics[width=0.8\columnwidth]{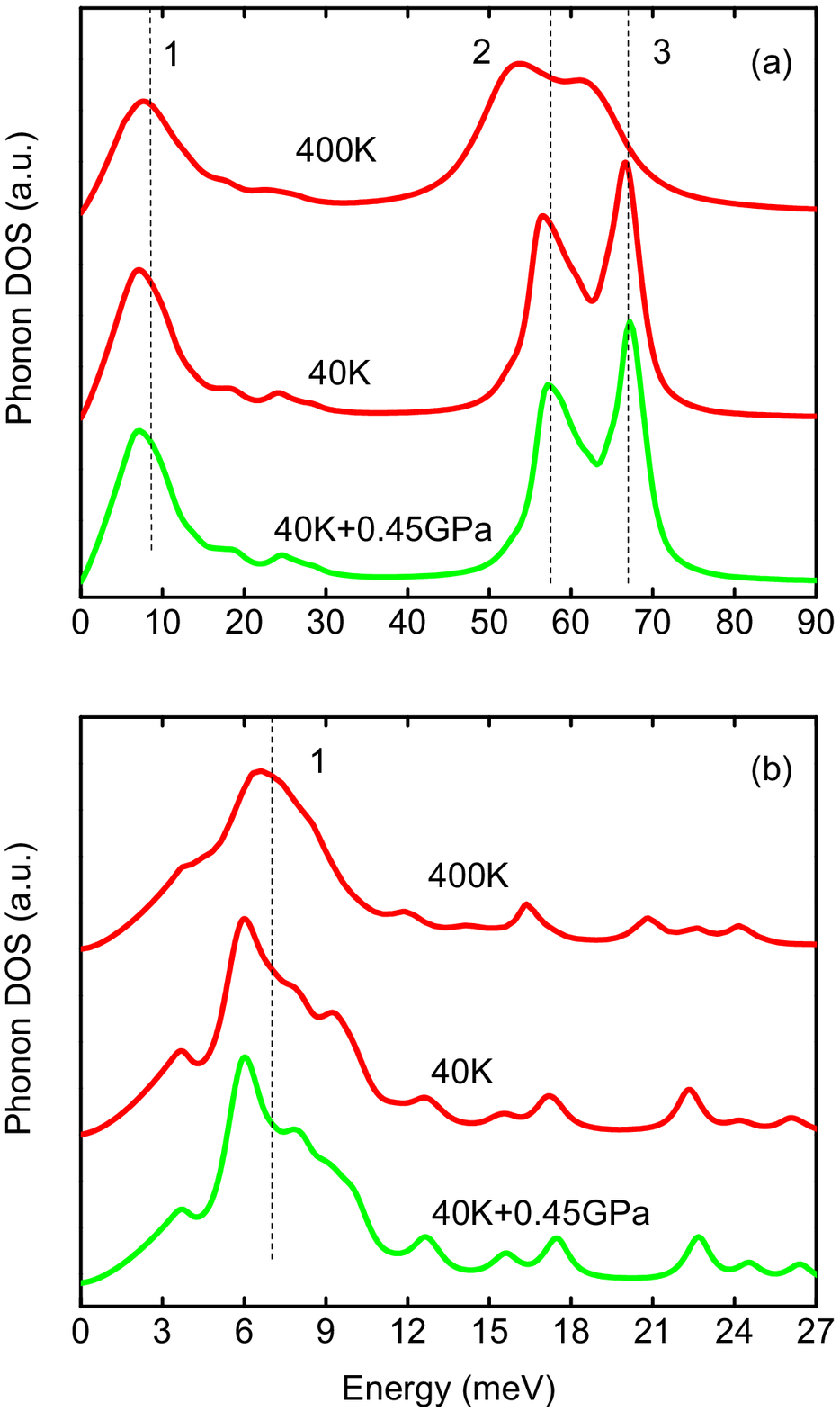}
\caption{Neutron weighted phonon DOS of Ag$_2$O with the cuprite structure from MD simulations. 
The green spectrum is the MD simulated phonon DOS at 40\,K and 0.45\,GPa. Vertical dashed lines are aligned to the major peak centers at 40\,K and labeled by numbers. 
The incident energy was 100\,meV for panel (a), and 30\,meV for panel (b). The spectra were convoluted with the resolution function characteristic of ARCS for the different energies of the incident neutron beam. }
\label{DOSSM}
\end{figure}

In a molecular dynamics simulation, the quasiharmonic contribution can be evaluated explicitly by turning off the temperature-dependent anharmonicity. In principle, this method is equivalent to the QHA method implemented with  self-consistent lattice dynamics,\cite{ScF3} 
and in practice we have found this to be true.
For example, we performed simulations at 40\,K for volumes characteristic of 400\,K, 
which produced a pressure of 0.45\,GPa. This calculation therefore removed the temperature effect while preserving the quasiharmonic volume effect at 400 K. The corresponding phonon DOS curves from MD calculations are shown in Fig.~\ref{DOSSM}. By comparing the phonon spectrum of a simulation at 40 K and 0.45 GPa with a simulation at 400 K, the pure temperature
dependence is identified.
From the spectra of Fig.~\ref{DOSSM}, it is found that the explicit anharmonicity dominates
the softening and broadening of the phonon spectra. 
All features in the phonon spectra with the QHA 
showed little change with temperature, except for small stiffenings at high energies. 

\bibliography{ref}